\documentclass[letterpaper,twocolumn,10pt]{article}
\usepackage{comment}
\usepackage{usenix2019_v3}

\usepackage{amsmath}
\usepackage{wrapfig}
\usepackage{colortbl}
\usepackage{url}
\usepackage{color,soul}
\usepackage{graphics}
\usepackage[inline]{enumitem}	
\usepackage{listings} 
\usepackage{booktabs}	
\usepackage{lipsum}
\usepackage{subcaption}
\usepackage{boldline}
\usepackage{amssymb} 
\usepackage{pifont} 
\usepackage{capt-of}	
\usepackage{todonotes}
\usepackage{setspace}

\definecolor{refkey}{rgb}{0,0,1}
\definecolor{labelkey}{rgb}{0,1,0}

\definecolor{airforceblue}{rgb}{0.36, 0.54, 0.66}
\definecolor{frenzyorange}{RGB}{249, 158, 26}
\newcommand*\circled[1]{\tikz[baseline=(char.base)]{
		\node[shape=circle,fill=frenzyorange,draw,text=white,inner sep=1pt] (char) {#1};}}
		
\renewcommand{\paragraph}[1]{\vskip 3pt\noindent\textbf{#1 }}	 
  {\begin{list}{$\bullet$}%
     {\setlength{\parsep}{0pt}%
      \setlength{\topsep}{0pt}%
      \setlength{\itemsep}{2pt}}}%
  {\end{list}}
\newcommand\note[1]{\sethlcolor{yellow} \hl{#1}} 
\newcommand\noted[1]{} 
\newcommand\lwg[1]{\sethlcolor{green} \hl{lwg: #1}} 

\newenvironment{myitemize}%
  {\begin{itemize}
	[leftmargin=0cm,itemindent=.3cm,labelwidth=\itemindent,
		labelsep=0pt,
		parsep=0pt,
		topsep=1pt,
		itemsep=2pt,
		align=left]
  }%
  {\end{itemize}}    

\newenvironment{myenumerate}%
  {\begin{enumerate}
	[leftmargin=0cm,itemindent=.5cm,labelwidth=\itemindent,
		labelsep=0pt,
		parsep=1pt,
		topsep=1pt,
		itemsep=3pt,
		align=left]
  }%
  {\end{enumerate}}

\newenvironment{enumerateinline}%
  {\begin{enumerate*}
	[label=\roman*)]
  }%
  {\end{enumerate*}}

\newcommand\sect[1]{Section~\ref{sec:#1}}	
        
\newcommand{\code}[1]{\texttt{#1}}

\newcommand{\fs}{file system}
\newcommand{\sys}{\code{Overwatch}}
\newcommand{\tz}{TrustZone}
\newcommand{\device}{smart device}

\newcommand{\fscore}{storage substrate}


\begin{document}
	
	
	
	
	
	%
	
	\title{Let the Cloud Watch Over Your IoT File Systems}	
	%
	%
	%
	%
	%
	
	%
	\author{
		%
		%
		Liwei Guo, Yiying Zhang, Felix Xiaozhu Lin\\
		Purdue ECE
	}


	\maketitle

%



\begin{abstract}

Smart devices produce security-sensitive data and keep them in on-device storage for persistence.
The current storage stack on smart devices, however, offers weak security guarantees: not only because the stack depends on a vulnerable commodity OS, but also because smart device deployment is known weak on security measures. 
To safeguard such data on smart devices, we present a novel storage stack architecture that 
i) protects file data in a trusted execution environment (TEE); 
ii) outsources \fs{} logic and metadata out of TEE; 
iii) running a metadata-only \fs{} replica in the cloud for continuously verifying the on-device \fs{} behaviors. 
To realize the architecture, we build \sys{}, a \tz{}-based storage stack. 
\sys{} addresses unique challenges including discerning metadata at fine grains, hiding network delays, and coping with cloud disconnection. 
On a suite of three real-world applications, \sys{} shows moderate security overheads.

\end{abstract}




\section{Introduction}
\label{sec:intro}

Smart devices, such as security cameras, voice assistants, and cleaning robots, emerge to be important cyber-physical systems. 
Unlike generic platforms such as PC, 
a smart device centers on a specific mission, e.g., capturing/analyzing videos or responding to voice commands. 
For engineering ease, they typically run commodity OSes such as Linux~\cite{rev-eng-wyze-cam,rev-eng-mi-robot,rev-eng-echo}.

\begin{figure}[t!]
	\centering
	\vspace{5mm}
	\includegraphics[width=0.48\textwidth{}]{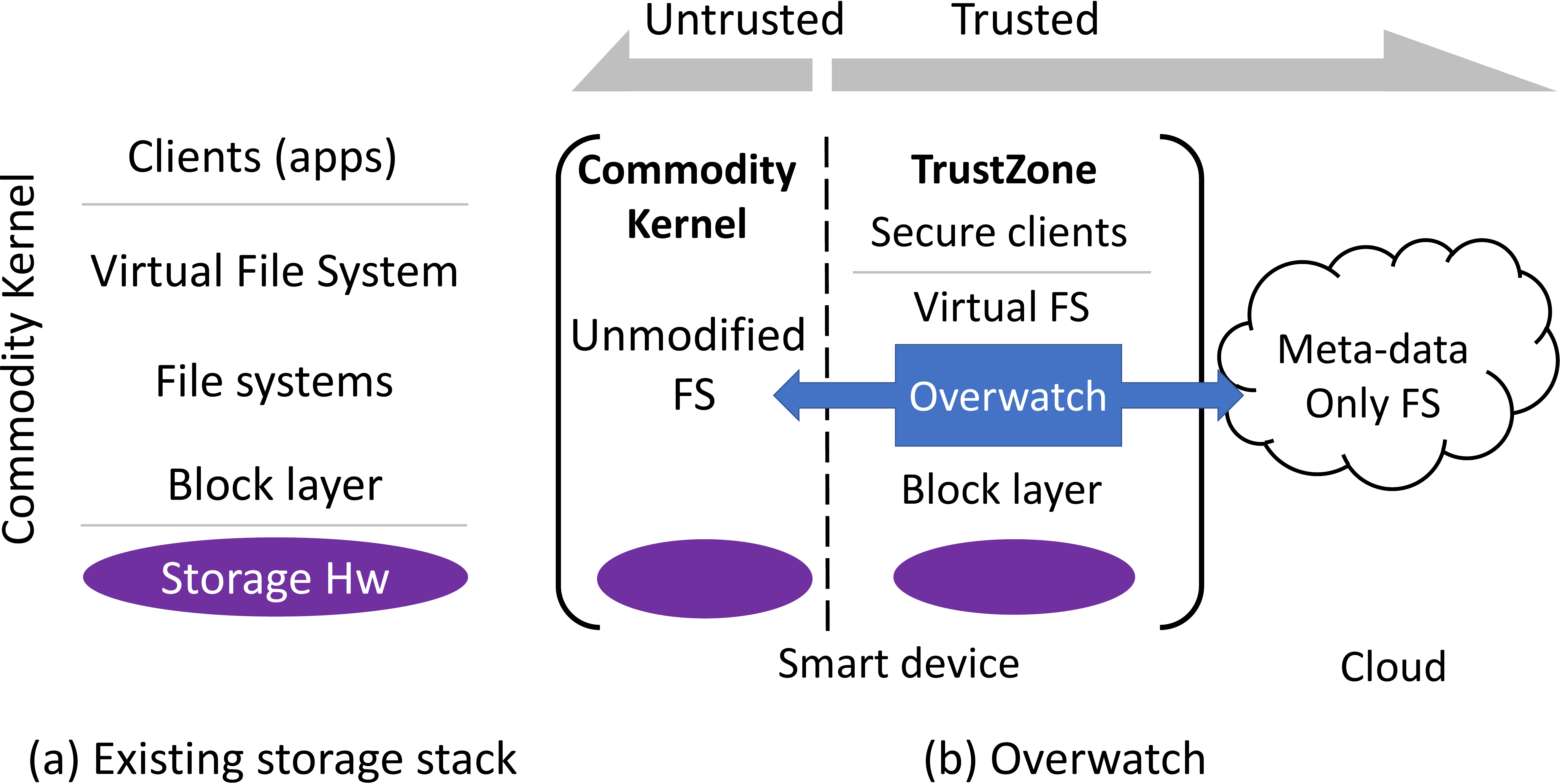}
	\caption{An overview of this work}
	\label{fig:overview}
\end{figure}

During operation, smart devices continuously generate data, e.g. video footage, floor maps, and flight logs. 
On one hand, the data is often confidential, e.g. for containing personally identifiable information; 
uploading the data to public cloud is often undesirable.
On the other hand, the data often has high business value, e.g. for containing important traffic events; 
data loss should be prevented.  
Smart devices, in the face of limited memory and possible power failures, 
often write the data to local non-volatile storage and may retrieve the data for processing later. 
In this process, the data is handled by a storage stack which spans file systems, the block layer, and storage hardware, as shown in Figure~\ref{fig:overview}(a).
It is the storage stack's responsibility to safeguard the data: not only keeping the data confidential but also assuring that the data has been correctly kept persistence and can be retrieved in the future. 


Unfortunately, the storage stack on today's smart devices is incapable of such guarantees. 
Vulnerabilities in \fs{}s and their runtime environment, a commodity OS kernel, are not uncommon. 
The threats are further amplified by smart devices' weak IT management, e.g. weak passwords and delayed security patches~\cite{ibm-iot-security-5facts,iot_gateway,iot-sec,iot-sec-trend}.
Attackers, on-device or remote, may exploit vulnerabilities through the user/kernel interface or the interface exposed by privileged network services and hence compromise the smart device OS kernel. 
They may learn the data, inject fabricated data, or delete data covertly. 

Prior solutions were inadequate in addressing these threats. 
Cryptographic file systems~\cite{cfs,ncryptfs,cryptfs} are subject to Iago attacks from a compromised kernel~\cite{iago-attack}. 
They can only detect data integrity violation \textit{in retrospect} which is less effective for memory-limited smart devices: upon the detection of violation, the data may already be lost permanently. 
File system checkers~\cite{sqck,recon} and kernel checkers~\cite{sprobes,skee} look for violation of  \textit{envelope} behaviors and also do so in retrospect. 
These checkers often demand substantial CPU/memory resource, likely unavailable on smart devices. 
Kinetic disks~\cite{pesos} push the check logic down to storage hardware; 
it requires new hardware support yet to be seen on off-the-shelf smart devices. 

 Modern CPUs offer Trusted Execution Environments (TEE) which are strongly isolated from the commodity kernel. 
ARM TrustZone supports a TEE to fully own physical memory regions and storage hardware, e.g. an eMMC RPMB partition~\cite{santos}.
While a TEE is already used for isolating program code that accesses security-sensitive data~\cite{trustshadow} and the underlying storage hardware, it sees difficulty in enclosing file systems, the largest portion of the storage stack.
On one hand, commodity file systems are diverse, feature-rich,
and have deep dependency on the kernel.
Porting them into TEE would bloat the TEE with substantial \fs{} and kernel code (as well as their vulnerabilities). 
On the other hand, 
reinventing new \fs{}s for a TEE, commonly seen in today's TEE-based systems~\cite{sgx-fs,optee-fs}, gives away mature features of commodity \fs{}s and fragments the \fs{} ecosystem. 

To achieve strong security properties while reusing existing \fs{}s as much as possible, 
we follow the principle of least privilege~\cite{least-privilege} and take an \textit{outsource and verify} approach: keeping any storage code out of TEE, as long as the code needs no access to data; verifying the outsourced code with a trusted party. 
This raises three primary questions. 

First, how to identify a clean, narrow boundary for outsourcing?
Rather than relying on 
sophisticated code analysis and slicing~\cite{glamdring}, 
our insight is that the boundary \textit{already} exists in a commodity kernel: 
all file systems export generic interfaces to the page cache above and to the block layer below (see Figure~\ref{fig:overview}(a)). 
The interfaces only contain several functions that work 
in a message-passing fashion. 
Hence, we partition the storage stack at these two interfaces and outsource \fs{}s in between. 
This makes the current \textit{functional} boundary a \textit{protection} boundary. 

Second, how can a \fs{} operate properly when it is strongly isolated from the underlying storage? 
Our insight is that file systems, in principle, operate only on \textit{metadata} but not file data. 
Hence, we keep file data inside the TEE while serving the \textit{metadata} to the \fs{} outside of the TEE.

Third, which party can be trusted for verifying the behaviors of outsourced \fs{}? 
Our insight is that the cloud, with rigorous security management, offer a more trustworthy execution environment than smart devices.
Therefore, we run a novel, metadata-only \fs{} replica in the cloud; 
the replica's sole goal is to validate the behaviors of the local \fs{}. 
This is shown in Figure~\ref{fig:overview}(b).
At run time, the TEE sends any invoked file APIs to a pair of \textit{twin} \fs{}s: the local, untrusted file system (``evil twin'') and an in-cloud, trusted replica (``good twin''). 
The TEE may perform storage operations advised by the evil twin (fast), but will only accept the outcome when the good twin confirms the operations (slow). 

The resultant advantages are threefold.
We offer strong confidentiality: file data never leaves the on-device TEE (not even to the cloud).
We bring the trustworthiness of the storage stack on smart devices to the level of its counterpart running in a rigorously-managed datacenter.
We reuse \textit{unmodified} commodity \fs{}s and only add light code to the TEE.

To make the approach practical, we address multiple unique challenges: 
ensuring file data to flow only in TEE, 
discerning metadata at fine grains,
hiding network delays, 
and continuing serving file access even when the cloud is disconnected. 
We build \sys{}, a concrete implementation based on ARM \tz{} with a suite of new designs. 
Our experiments show that \sys{}'s security mechanism incurs moderate overhead in representative smart device applications: 
15\%-45\% increase in application latency and 5\% loss in application throughput.

This paper makes the following contributions:
\begin{myitemize}
\item 
Towards securing storage on smart devices, we conduct an analysis of their file IO, the attacks they face, and how existing solutions fall short in defeating attacks. 

\item 
We present a secure storage stack architecture for smart devices. 
It outsources file system logic and metadata while protecting file data. 
It runs a novel, metadata-only \fs{} replica in the cloud for continuous verification. 
We analyze how the new architecture thwarts the aforementioned attacks.

\item 
We build \sys{}, a concrete secure storage stack that incarnates the proposed architecture. 
\sys{} addresses system challenges with a suite of novel designs, including secure data path, metadata stencils, and emergency files. 

\item
We demonstrate that \sys{} works with ext2 and f2fs, two popular, unmodified \fs{}s.
Atop \sys{}, we build multiple real-world smart device applications which show moderate security overhead. 

\end{myitemize}

\vspace{-3mm}
\section{Background \& Motivations}
\label{sec:background}
\label{sec:bkgnd}
\label{sec:motiv}


\subsection{\tz{} and its unique features}
\label{sec:tee}

A Trusted Execution Environment (TEE) is isolated by hardware, as exemplified by Intel SGX~\cite{intel-sgx} and ARM \tz{}~\cite{arm-trustzone}.
Unlike SGX, \tz{} partitions all hardware resources of a System-on-Chip (SoC) into a normal (insecure) world and a secure world.
In particular, the security of our system benefits from the following \tz{} features: 

\paragraph{1) Physical memory partitioning.}
Unlike SGX where the untrusted OS maps memory pages to the TEE dynamically \textit{at runtime}, 
\tz{} partitions the physical memory statically at \textit{boot time}. 
The normal world is strongly isolated from the secure memory and therefore cannot mount controlled-channel attacks~\cite{controlledChannel} against the latter.

\paragraph{2) IO partitioning.}
Unlike SGX where TEE accesses IO through the untrusted OS, 
\tz{} isolates peripherals by statically assigning them to different worlds at boot time.
It does so by assigning IO interrupts and IO memory regions through the \tz{} Protection Controller.
This allows the secure world to fully own on-device storage hardware, e.g. one SD card or a specific partition on an eMMC device~\cite{santos}.
We refer to such storage devices owned by TEE as secure disks\footnote{While recognizing that smart devices often use flash-based storage, we use ``disk'' as a generic term for non-volatile storage} in the remainder of this paper. 






\subsection{The storage stack}
We use Linux, one of the most popular OSes for smart devices~\cite{motioneyeos,rev-eng-wyze-cam,rev-eng-mi-robot,rev-eng-echo}, as the example. 
The storage stack is illustrated in Figure~\ref{fig:overview}(a).
At its top, the stack provides a \textit{Virtual File System} (\textit{VFS}) layer, which caches recent file data (via a page cache) and exports a set of common APIs for all concrete file systems (e.g. ext3 or f2fs) to implement. 
All \fs{}s invoke a common block layer, which serves block I/O requests from the former and accordingly drives the disk (through device drivers) below. 

In response to a client's file API invocation (e.g. ``read 42 bytes from /local/data at offset 100''), VFS first attempts to serve the invocation from its page cache. 
Upon cache miss, VFS invokes the underlying file system, which translates the API invocation to disk block operations (e.g. ``read from block 21''). 
The block operations are served by the block layer. 
In this process, the file system inspects metadata, e.g. inode table, for which it may trigger additional block operations. 
At the end, the kernel copies the read block data to the page cache and then the client's buffer. 
File write is a mirror process. 

\subsection{File IO patterns on smart devices} 
\label{sec:motiv:observation}

Unlike general-purpose platforms such as servers and PCs, smart devices are single-purpose, e.g. security cameras for surveillance, 
robots for cleaning floors, 
voice assistants for responding to user commands.
During operation, a smart device generates data pertaining to user interest or its own operation. 
Our study of multiple off-the-shelf smart devices (see Section~\ref{sec:impl} for details) reveals
the following IO patterns which have strong implications on our security objectives. 

\paragraph{User data is the focus of protection}
We use ``user data'' to refer to the data produced by smart device  operations. 
Examples include captured videos, learnt user voice models, and robot operation logs. 
User data is often privacy-critical, e.g. user models may be tracked back to individuals;
the data is often business-critical, e.g. captured videos may contain important crimes and accidents. 
It is crucial to prevent user data leakage and loss.

Besides user data, smart devices are preloaded with system data, e.g. program binaries or configuration files, often located in directories different from user data. 
System data has lower value for protection: 
one could dump the system data from any off-the-shelf smart device of the same model.



\paragraph{Directory structures are pre-defined}
On smart devices, the structures of file directories, including the tree topology, the numbers of subdirectories at each tree level, and the numbers of files in each subdirectories, are typically pre-defined at device development time. 
They cater to the device's application logic and often remain unaffected by user data discussed above. 
One could learn the pre-defined directory structure by dumping the storage of off-the-shelf devices~\cite{rev-eng-echo,rev-eng-mi-robot}. 

We find this pattern in most, if not all, popular smart devices. 
\begin{enumerateinline}
\item  
Security camera systems, including WyzeCam~\cite{rev-eng-wyze-cam} and MotioneyeOS~\cite{motioneyeos}, store captured videos as same-length footage, under directories organized by time ranges. 

\item 
Voice assistants, such as the opensource Mycroft~\cite{mycroft}, keep each user-defined rule (``skill'') and the corresponding response in a dedicated directory; 
while operating, it stores captured voices as same-length clips under /tmp. 
Amazon Echo Dot, a commodity voice assistant, is likely to have similar behaviors based on the limited information revealed in reverse engineering~\cite{rev-eng-echo}. 

\item
Robot cleaners, such as Xiaomi Mi, stores all the operation logs as well as the floor map in one specific directory~\cite{rev-eng-mi-robot}.

\item
Drones, such as DJI Phantom 2, stores all sensor data (e.g. images with EXIF data) in a directory structure similar to security cameras~\cite{rev-eng-dji-phantom2}.
\end{enumerateinline}

\paragraph{Block access patterns are regular}
Being single-purpose, smart devices show regular block-access patterns as driven by their application logic, which again is orthogonal to user data. 

Regardless of captured video contents,
a security camera creates new directories periodically and keeps appending data to video files. 
During one operation, the cleaning robot keeps appending to log files; after the operation, it reads back log files sequentially and writes to a floor map file sequentially. 
A voice assistant periodically scans its ``skill files'' and write recorded audio samples to wav files sequentially. 

Such regular access of file contents, combined with access to pre-defined directory structures, result in regular, repetitive block traces, e.g. querying the super block, read one metadata block and a fixed number of data blocks, and write one metadata block back.


\paragraph{Implications}
On smart devices, the user data resulted from device operation should be the focus of protection. 
By contrast, the system data, the directory structures, and the block access pattern, are pre-defined, can be learnt by analyzing off-the-shelf devices, and therefore have lower security values. 


\subsection{Security threats \& design objectives}
\label{sec:motiv:threats}
Smart devices suffer from common weakness of IoT, notably weak passwords and delayed security patches~\cite{symantec-iot-security, windriver-iot-security}.
Local and remote adversaries, through compromising the smart device OS, can break the security of user data. 

Example attack paths are as follows. 
i) A local unprivileged adversary may exploit \fs{} bugs to corrupt the file data~\cite{CVE-2009-4306,CVE-2009-4131}.
ii) Local unprivileged adversaries may exploit kernel vulnerabilities through the user/kernel interface~\cite{CVE-2018-9422}. 
iii) Remote adversaries may exploit may exploit vulnerabilities in privileged network services (e.g. an HTTP server~\cite{CVE-2014-0226}) or the kernel network stack~\cite{CVE-2016-7117}.
For ii) and iii), a successful adversary either gain the root privilege or become capable of executing arbitrary code in the kernel context.
She then replaces key functions in the file system, e.g. \code{submit\_bh()} which moves data to/from the block layer.
Her own malicious substitute for the function may to reveal, modify, or drop the user data in the file system.



\paragraph{Observation: cloud as a more trustworthy environment}
Like smart devices, the servers in datacenters also face threats from the
vulnerabilities in their system software. 
Unlike smart devices, 
the servers are hosted in a more trustworthy environment. 
i) Datacenters follow rigorous standards~\cite{nist} and mature protocols~\cite{googlecloud} in the face of incidents and vulnerability. 
By contrast, smart devices are often configured in batches and weakly~\cite{verizon_2017};
their much delayed security patches leave a large window of attacks. 
ii) Datacenters can afford heavyweight security measures including  frequent kernel introspection and regular file system checkers. 
By contrast, these measures are often unaffordable to a smart device, which only has a few CPU cores, a few GB DRAM, and limited power supply.

Fortunately, today's smart devices are typically designed under the assumption of cloud connectivity for enriching their functionalities.
This observation motivates us to secure the smart device storage with the assistance of the cloud. 

\paragraph{Objectives} 
\textit{Strong confidentiality for file data.} 
We ensure that user data never leaves the on-device TEE. 
Catering to smart device file IOs (\S\ref{sec:motiv:observation}), we carefully choose to protect file contents, file names, and directory names; 
we do out protect directory structures and file system metadata including super blocks, allocation maps, and inodes; 
we do not protect block access patterns which are regular.  

\textit{Continuous assurance of data integrity and persistence}.
We set to bring the trustworthiness of a storage stack running on smart devices to the level of its counterpart running in a rigorously-managd datacenter. 
At this level of trustworthiness, the storage stack ensures that: 
a successful read reflects the most recent write to the same file location; 
a successful \code{fsync()} implies that the data becomes persistent on the disk. 

\textit{Practicability.} 
We set to respect the diverse, mature \fs{}s by reusing their code with little modification.
We set to add as little code to TEE as possible; 
we set to keep the interface exposed by TEE as narrow as possible.

\section{Security approach overview}
\label{sec:overview}

\subsection{Scope}

\paragraph{Target scenarios}
We target smart devices for recording/analyzing environment data and/or serving human users; 
such devices are commonly seen in homes or offices, as exemplified by security cameras and voice assistants. 
We recognize the significance of mission-critical devices with tight control loops, but do not target it.

During operation, the smart devices generate data (``user data'') that is privacy-critical and/or business-critical; the smart devices store the data to files for persistence. 
We trust the \tz{}-based TEE on a smart device; 
the TEE already encloses a secure disk as well as app code (i.e. secure clients) that accesses the protected user data. 
We assume cloud connectivity but nevertheless design to cope with poor connectivity or even disconnection. 
We assume a ``honest but curious'' cloud service that execute timely patched, thoroughly checked file systems; 
however, we do not trust the cloud for data confidentiality.

\paragraph{In-scope threats.}
We consider malicious adversaries interested in learning user data and tampering with it. 
We assume powerful adversaries: it takes full control of the smart device's OS, including the encompassed file systems, network stack, and any user processes atop the OS.
The compromised file system may alter or covertly drop write requests; 
it may supply wrong data to read requests.

\paragraph{Out-of-scope threats.}
We consider the following threats out of scope: 
\begin{enumerateinline}

\item 
Exploitation of TEE kernel bugs~\cite{vtz, CVE-2015-4421, CVE-2015-4422}. 

\item
Physical attacks, e.g. snooping TEE's DRAM access~\cite{tampering,lowCostAttack}. 

\item
Availability attacks, e.g. a compromised OS could refuse to boot or deny requests from TEE; 

\end{enumerateinline}
Many of these attacks are mitigated by prior work~\cite{sel4,opaque,sentry,TZ-CaSE} orthogonal to \sys{}. 
Note that controlled-channel attack~\cite{controlledChannel} does not apply to ARM TrustZone as the latter's page management is within TEE unlike Intel SGX. 





\subsection{Existing approaches are inadequate}
\begin{table}[t!]
	\fontsize{8.0}{9.0}\selectfont
	\begin{center}
		\setlength\tabcolsep{3pt}	
		\begin{tabular}{
				l	
				l	
				l 	
				c	
				c	
			}
			\hlineB{2.5}
			\cellcolor{lightgray}\textbf{System} &
			\cellcolor{lightgray}\textbf{TCB} & 
			\cellcolor{lightgray}\textbf{SG}
			\\ 
			\hlineB{1.5}
Cryptographic FS~\cite{cfs,ncryptfs,scfs} 				 	
& OS
& \texttt{CI-} \\ 


FS checkers~\cite{sqck,recon}
& OS
& \texttt{-I-} \\

Kernel checkers~\cite{sprobes,skee}
&  TEE
& \texttt{C$--$} \\

Outsource w/o verify~\cite{scone,graphene-sgx,obliviate,trustshadow}
& TEE
& \texttt{CI-} \\


\sys{} (this work)
& TEE
& \texttt{CIP} \\

\hlineB{2.5}
		\end{tabular}
		\fontsize{7.3}{7.7}\selectfont \\
		\textbf{SG}: security guarantees. \\ 
		\texttt{C}: data confidentiality; 
		\texttt{I}: data integrity; 
		\texttt{P}: persistence assurance\\
		\caption{A comparison of existing solutions and this work}
		\label{tab:cmp}
	\end{center}
\vspace{-2mm}
\end{table}

Cryptographic file systems \cite{cfs,plutus,sirius} guarantee data confidentiality/integrity 
but not file system correctness;
they are also subject to Iago attacks~\cite{iago-attack} from a compromised kernel, e.g. by overwriting data blocks and hence causing permanent data loss.  
To enforce an \textit{envelope} of file system behaviors, a file system can be certified through formal methods~\cite{fscq,sibylfs} or checked at run time~\cite{inktag}.
However, most commodity file systems are \textit{not} built with formal methods; 
envelopes does not capture all possible file system behaviors;
deploying per-file-system checkers into the smart device TEE is likely to increase the edge TCB and the overhead significantly~\cite{recon, sqck}.
Data auditing~\cite{pdp,provenance-for-cloud} proves data possession but not \textit{persistence}.
Non-repudiable IO~\cite{non-repudiableio} ensures that given disk reads or writes hit the disk while not asserting file system correctness.
While much prior work can detect damicrobenchmarkta loss in retrospect,
few techniques prevent it from happening.

To support TEE code for accessing storage, existing TEE-based systems take ad-hoc solutions: 
leaving the whole storage stack out of TEE and delegating file APIs to it~\cite{sgx-fs,scone,optee}, 
pulling an entire storage stack to TEE, 
or hand-crafting a miniature stack (e.g. by only supporting a set of predefined files)~\cite{obliviate,trustshadow,ryoan}. 
Lacking systematic treatment, they suffer from Iago attacks, bloating TEE, and giving up decades of file system development, respectively. 
As \fs{}s are tightly coupled with the kernel environment, 
pulling a whole \fs{} into TEE would end up pulling most, if not all, of the kernel dependency (at least 30K SLoC): address space and page management (15K), memory allocator (7K), locking (8K), etc. 
This bloats the TEE codebase. 

\begin{figure}
	\centering
	\includegraphics[width=0.46\textwidth{}]{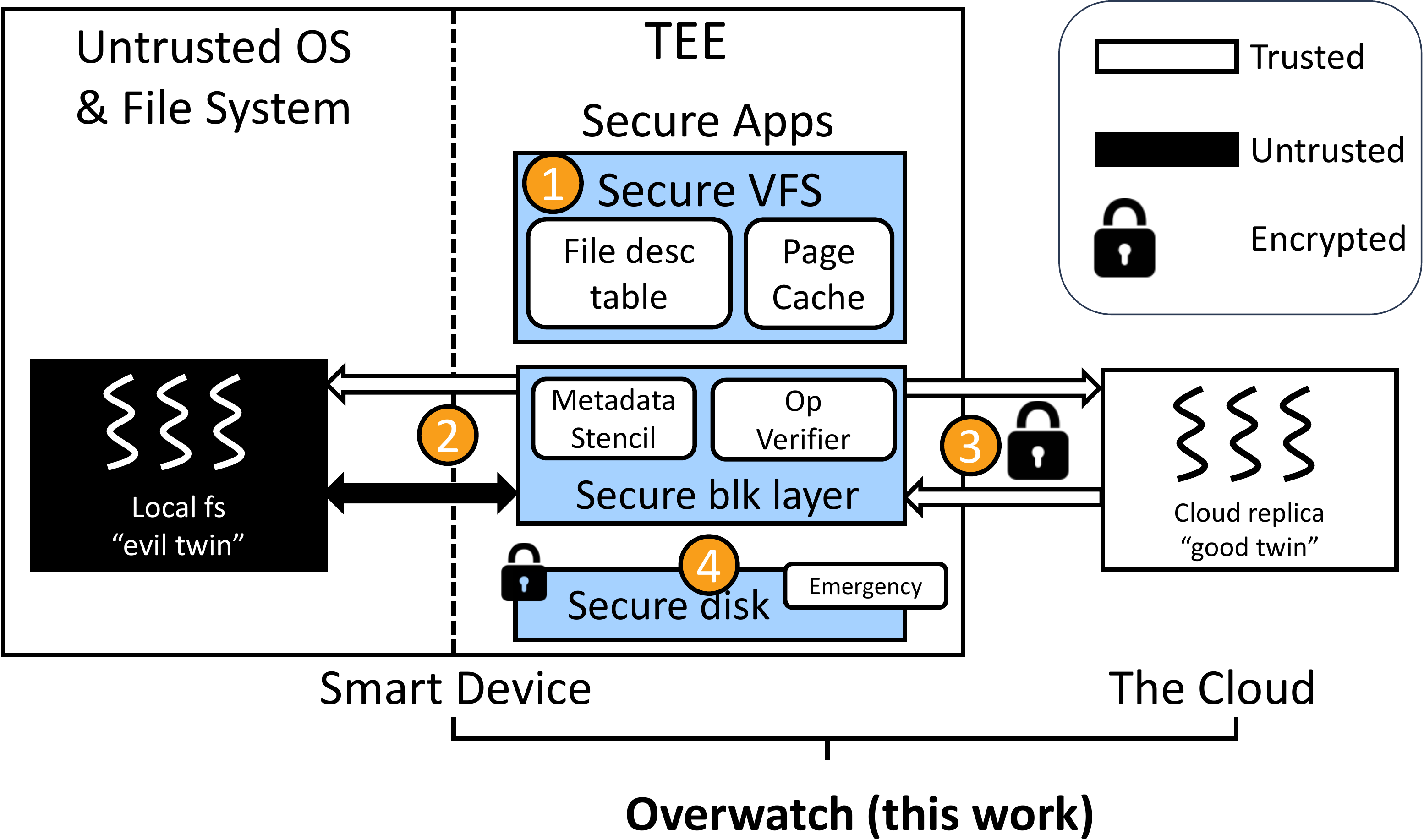}
	\caption{The \sys{} architecture}
	\label{fig:arch}
\end{figure}

\subsection{Our key ideas}
\label{sec:overview:idea}


Our observations are:
i) although the \fs{}s are complex, the data functions, i.e. page cache and block layer, are much simpler and hence
fit TEE; 
ii) although file systems have disparate internals, they operate on storage data through a narrow, unified block interface; 
iii) compared to smart devices, the cloud offers a more trustworthy environment to file systems (\S\ref{sec:motiv:threats}).

Accordingly, we propose the following designs as illustrated in Figure~\ref{fig:arch}. 

\begin{myenumerate}
\item 
\textit{Protecting file data while outsourcing file system logic.} 
We partition the current storage stack: within TEE we isolate VFS, which tracks opened files and serves page cache, and the block layer, which guards a trusted disk\footnote{While recognizing that edge platforms often use flash-based storage, we use “disk” as a generic term for non-volatile storage} owned by TEE and all the data on the disk.
We leave the unmodified code of commodity file systems (e.g. ext2) in the insecure world and consider the code untrusted (i.e. the ``evil twin''). 

\item 
\textit{Exposing metadata to the untrusted local file system.} 
As the file system code needs to operate on the metadata (e.g. file
system inode table, block bitmap, and directory structure), the storage substrate exposes an interface for the insecure world to access the metadata. 

\item 
\textit{Cloud as the verifier}. 
We run a trusted, lightweight replica of the same file system, i.e. the ``good twin'', in the cloud. 
Its only responsibility is to validate the legitimacy of any storage operations that the untrusted edge file system suggests to execute. 
The cloud replica timely incorporates newest bugfixes, but no new file system features.
\end{myenumerate}

In this architecture, only the device TEE possesses the file data; both the device's insecure world and the cloud work on their own metadata copies.

%
%

\paragraph{Workflow}
As shown in Figure~\ref{fig:arch}, a secure client invokes file API (e.g. ``write to file /local/data offset 42''). 
If the data happens to be in the secure page cache (\circled{1}), the execution never leaves the TEE for consulting \fs{}s. 
Upon cache miss, \sys{} sends the invocation to both the local \fs{} (the untrusted evil twin) and the cloud replica (the trusted good twin), which both return storage operations resulted from the invoked API (e.g. ``writing the given data [opaque reference] to disk block 42'') (\circled{2}\circled{3}).
Conceptually, only when the twins return the same operations, the secure block layer executes such \textit{validated} operations on the protected storage (\circled{4}).
With this architecture, the good twin (remote) ultimately guarantees the security objectives, 
while the evil twin (local) is crucial to performance optimization, as will be discussed in \sect{design}.

\paragraph{Security benefits}
i) Our approach offers strong confidentiality over the file data. 
By exploiting \tz{}'s static partitioning of memory and IO hardware (\S\ref{sec:bkgnd}), 
we ensure file data, 
as well as the memory and storage hardware containing the data, 
are strongly isolated from the rest of a smart device, including the commodity OS. 
The autonomy of \tz{} eliminates controlled channel attacks through page faults~\cite{obliviate}.
The file data is never disclosed to the cloud either, as the latter only operates on metadata. 
ii) Our approach offers high assurance of correctness: 
the assurance no longer depends on the integrity of local file system and OS that face high threats, but on the integrity of the cloud. 
By continuously validating the block-level activities at run time, 
it assures the apps that file data is safely kept persistent and can be readily read back.

\section{\sys{} Design}
\label{sec:design}

Our security approach above takes a somewhat idealistic position. 
To realize the model and make it practical, we have addressed the following challenges with novel system designs. 

\begin{myenumerate}
\item How to make file data flow between the clients and the storage hardware without leaving TEE?
\item How should the TEE differentiate file data and metadata? 
\item How to hide long network delays? 
\item How to ensure consistency between twin \fs{}s? 
\item How to continue operating when the cloud is disconnected? 
\end{myenumerate}

\subsection{Secure data path}

As discussed in Section \ref{sec:overview:idea}, \sys{} incorporates in TEE the VFS and the block layer from the commodity OS. 
These layers are generic; compared to \fs{}s, they are thinner, only adding 1K SLoC to the TEE code. 

It has relatively simple responsibilities: manage file-related data structures (e.g., file descriptor table and page cache), and call into the \fs{}-specific control functions.
Hence, \sys{} also implements a simple yet generic virtual file system layer, \sys{} secure VFS~(\circled{1}).
It provides a basic abstraction for opened files through a secure file descriptor table, and the protection of data through secure data path.

\paragraph{Secure file descriptor table} 
To secure clients \sys{} presents POSIX file APIs. 
To do so, it keeps in TEE a file descriptor table, keeping track all opened files and their current access positions.  
When forwarding file APIs to the local \fs{}, \sys{} obfuscates all file names and bookkeeps the mapping between the obfuscated file names exposed to the \fs{} and the actual file names used by secure clients. 

\paragraph{Secure page cache} 
\label{design:cache}
is where the \sys{} holds the recently accessed file data in memory. 
Similar to Linux's page cache, the secure page cache is essentially a dictionary keeping $\langle file, offset \rangle$ $\rightarrow$ page, block\_id. 
Any file API hits the cache will be served by the cache within TEE. 
Only upon a cache miss the \sys{} forwards the file API to the twins of file systems. 
The secure page cache only keeps user blocks not metadata blocks, which are manipulated by the local \fs{} with untrusted contents. 

And upon the delegated file operation returns, the verified blocks will be read from secure storage to fill corresponding pages.
This is because the secure world never relies on the metadata to perform control path.

\paragraph{Secure block layer}
The block layer exposes a very narrow interface: copy data between a TEE memory address and a block on the secure storage. 
For read(), the block layer copies a disk block to the secure page cache, and then to a buffer supplied by the secure app;
for write(), the block layer copies a secure app's buffer to the secure page cache; 
for sync(), the block layer copies data from secure page cache to disk blocks.

\subsection{Metadata stencils}
\label{sec:design:stencil}
Based on our idea of only exposing metadata (\S\ref{sec:overview:idea}), the TEE serves metadata to the untrusted file system code (Figure~\ref{fig:arch} \circled{2}). 
In doing so, \sys{} must i) reject any request to \textit{file} data;
ii) redact the metadata that contain user information (e.g. directory name) before serving;
iii) serve metadata in cleartext if it contains no user information.
It is worth noting the metadata includes directory links (but not directory names), which is required by the \fs{} for walking file paths. 
To do so, \sys{} must know metadata's disk locations, for which it relies on \textit{metadata stencils}.

\paragraph{What are metadata stencils?}
Metadata stencils, a compact data structure in TEE, encode the knowledge of metadata's disk locations. 
Conceptually, for each block on the secure storage, a stencil specifies the bytes that represent metadata, e.g. ``block 42: byte 0-127 [metadata]''.
To serve a block read from the local \fs{}, \sys{} consults the corresponding metadata stencil, reveals the metadata bytes, and redacts the remaining. 
similarly, for a block write \sys{} only overwrites the data bytes. 
In practice, the metadata stencils are highly compact:
file systems such as ext2/3 use separate disk blocks for metadata and data, allowing a one-bit stencil for each block; 
file systems such as BTRFS and F2FS may co-locate metadata and data in the same block, yet we find only such blocks only constitute a small fraction. 

\paragraph{How to generate metadata stencils?}
The metadata stencils are generated by parsing two on-disk data structures: the super block and inodes. 
The super block describes the usage of all live disk blocks; 
the inode structure describes any file data that may be embedded in an inode block. 
The two data structures are \fs{} specific; 
their layouts are stable across different versions of the same \fs{}, because a \fs{} typically keeps its disk layout backward-compatible. 
Upon \fs{} boots, a simple parser locates and parses the super block (identified by its well-known magic number) and inodes, establishing metadata stencils for all live blocks. 


\paragraph{Who should generate the stencils?}
We have investigated the following two solutions.
The cloud replica generates the stencils based on the metadata it possesses; it piggybacks the stencils to its validation responses. 
\sys{} runs on device without any \fs{}-specific logic;
it remains agnostic to the type of \fs{} (e.g. ext2 or f2fs) running out of TEE. 
\sys{} can be deployed to the TEE once and remains \textit{sealed} afterwards. 

Alternatively, \sys{} may generates and maintains the stencils by itself. 
This requires \sys{} to incorporate a separate parser for each type of \fs{}s it works with. 
Yet, data confidentiality is stronger: 
it completely depends on the TEE and independent of the cloud. 
We will test this solution in evaluation (\S\ref{sec:eval}). 

\subsection{Memoizing verified file-block mapping}
\label{sec:design:memoiz}
Unlike \code{write()}, an uncached and \textit{trusted} \code{read()} is synchronous.
Because \sys{} will have to wait for the cloud to return validated blocks to proceed, it inevitably incurs latencies that worsens linearly with elongated network delays and suffers much from unstable network connections.
To optimize for such cases, our key insight is to reduce the remote validation by memoizing the file offsets with their corresponding \textit{verified} data blocks and return the data blocks directly to the application.

Our rationale is that, the mappings from file offsets (i.e. the input of file system execution) to locations in page caches and disk blocks (i.e. the output of file system execution) will remain valid until future file API invocations alter the file system metadata.

It works like a cushion to the secure page cache: whenever a page in the secure page cache gets evicted due to limited memory size, \sys{} memoizes the blocks and file offset that map to the spilled page.
And after \code{read()} misses the search in page cache, it will then search among the memoized disk blocks and return the trusted data blocks if it finds the corresponding blocks.
Also, it implies that the \code{read} from the offset within the same disk blocks will not require remote validation, either.
For instance, in serving two consecutive reads at the same file at offset 42 and 128, the \sys{} avoids invoking the file system and the cloud for the second read, because the two offsets are known to map to the same block and the mapping is unchanged.
We will evaluate this mechanism in Section \ref{sec:microbenchmark}.

\subsection{Hiding network delays}
\label{sec:design:network}

\sys{} incorporates the following mechanisms. 
Essentially, it exploits the untrusted local file system for \textit{speed} and relies on the trusted cloud replica for \textit{correctness}.

\paragraph{Overlapping network/storage delays}
\sys{} executes storage operations from the (untrusted) local \fs{} as soon as they become available; the \sys{} rolls back the operations if it receives dissension from the cloud later.
By doing so, the \sys{} overlaps the network latency (typically tens to a few hundred ms) with the local storage latency (typically a few to tens of ms).

\paragraph{Overlapping network/computation delays}
As described before, \sys{} provides POSIX file APIs to its clients. 
It further provides an \textit{option} for the clients to observe the outcome of unvalidated storage operations.

Our rationale is to give the clients opportunities to handle unvalidated (and hence untrusted) file data with their own logic.
For instance, the camera code may process unvalidated images it reads from storage while the validation is still pending.
For \code{read()}, \sys{} returns the requested data to the client and indicates the data is yet to be validated;
for \code{write()},
\sys{} checkpoints the blocks to be modified (in case of future rollbacks) and taints the modified blocks and pages as untrusted (in case of future read from them).
In case the cloud rejects the data operations, \sys{} rolls back to the earlier version without the modifications.

To support the option, \sys{} introduces a light interface augmentation, by adding two flags to existing file APIs. 
In \code{open()}, \sys{} supports a new \textit{untrusted} flag, indicating that any subsequent access of this particular file will return with the results from the local \fs{} without waiting for the remote validation to complete. 
In \code{select()}, \sys{} supports a new \textit{validation} flag which serves as a validation barrier. 
\sys{} will block the caller client until all pending validations pertaining to this file are completed. 
The new flags are simple yet sufficiently powerful to support overlapping between computation and network delay. 
\sect{eval} will present a case study. 

\subsection{Coping with emergencies}
\label{sec:design:consistency}

\paragraph{Emergency file for cloud disconnection}~\sys{} supports clients to write a fixed amount of data \textit{reliably} without any remote validation.
This is important for keeping time-critical user data persistent, e.g. an image frame containing a person's face of interest.
To do so, during file system initialization and while the cloud is connected,
\sys{} creates an \textit{emergency} file and pre-allocates all its blocks and remembers all its data blocks by memoization, as described in Section~\ref{sec:design:memoiz}.
As the mappings between file offsets and disk blocks are all known to the \fscore{}, it can safely access the file without consulting the cloud.
The size of the emergency file is configurable to the device user.
\paragraph{Maintaining crash consistency}
\sys{} applies 2-phase commit (2PC) protocol~\cite{2pc} to maintain consistency between the two \fs{}s on the edge and the cloud.
On the local \device{}, for on each file operation that changes the metadata (e.g. \code{write} to a new file) \sys{} piggybacks an additional \textit{commit} request to the cloud;
on the cloud, when \sys{} receiving the file operation and the commit request, \sys{} executes the file operation but saves the modified metadata into a temporary file and piggybacks an \textit{ok to commit} message with the block requests.
Then after receiving the message, the \device{} executes the verified block requests and sends the final \textit{commit} message to the cloud. 
Lastly, only upon receiving the final commit from the edge should the cloud write the modified metadata persistently to its storage.
In this way, if the device crashes (e.g. due to power failure) before executing the file operations, the cloud will not receive the final \textit{commit} and thus will not update its copy of metadata; however, if it's due to temporary disconnection, on the next successful connection, the \device{} will resend the \textit{commit}, thus allowing the cloud to update its copy, keeping both copies consistent.

\subsection{Maintaining the cloud replica}
\label{design:ofs-cloud}
\paragraph{A metadata-only \fs{}}
The cloud replica of local device's \fs{}, called \sys{} \fs{} (\code{OFS}), runs in the cloud to verify the block requests of the untrusted \fs{}s on the \device{}. 
\code{OFS} has the same control logic as the local file system but operates only on the metadata
(i.e., inode table, block bitmap, inode bitmap, etc. in the ext family).
Its initial metadata is replicated when the local file system is first initialized (e.g., when \code{mkfs}).
Since then, the subsequent replication of metadata is performed by replaying each file operation sent by the smart device, where the consistency is maintained by 2PC protocol, as discussed in Section~\ref{sec:design:consistency}. 
It is worth noting that \code{OFS} only maintains and operates on its replica of metadata.
It never possesses or touches the actual user data.
Therefore, the replay is cheap and that the device is free from transferring its data to the cloud, which is worry-free of any potential user data leakage, and reduces the data transfer overhead aggressively, as will be shown in \S\ref{sec:macrobenchmark}.

\paragraph{Operation verifier} After sending the file operations to the cloud, the in-TEE operation verifier (shown in Figure~\ref{fig:arch}) receives the block requests generated by \code{OFS} and verifies those issued by the untrusted OS.
As discussed, \sys{} delegates file operations to both the untrusted \fs{} and the cloud replica (\code{OFS}).
Upon receiving the file operation, \code{OFS} executes the file operation by consulting and manipulating its replica of metadata;
as a result, it generates a sequence of block requests, and responds to the operation verifier with these block requests.
The operation verifier compares generated block requests.
The matching block requests (i.e. both the requested block number and request type are the same) result in a successful data read/write while failure leads to data roll back, as discussed in Section~\ref{sec:design:network}.

\section{Implementation}
\label{sec:impl}

We have built \sys{} in 5K lines of \code{C} code (reported by Sloccount~\cite{sloccount}). 
We implement the following major \sys{} components, which are agnostic to specific \fs{}s.

1) The \sys{} runtime within on-device TEE.
We build it atop OP-TEE OS v2.6~\cite{optee}, where we add 3K for new implementation and reuse the exception handling and RPC facility to handle world switch.
We emulate the secure storage as a ramdisk.
We implement the metadata stencil by porting only 300 lines of code from \code{libext2} and \code{libf2fs}. 
Block requests from the untrusted \fs{} are verified sequentially. 

2) A small kernel module in the smart device kernel (Linux v4.9) to forward the delegated file operations to the twin of \fs{}s.
It sets up shared memory as the secure communication channel, and establishes the network connection with the cloud.
To support multiple commodity \fs{}s, we also modify loop device driver of the on-device Linux kernel.

3) A secure communication channel between the TEE and the local \fs{}. 
The channel passes messages through shared memory which has a fixed size of 4KB.
The channel serves two purposes:
1) passing file APIs and block requests.
Invocations to file APIs are identified by \code{<op,fd,flag,count,name>}.
Block requests, resulted from untrusted \fs{}'s execution, are a sequence of block numbers represented by 32-bit integers. 
2) passing metadata blocks.
\sys{} serves metadata block requests after they are checked against the metadata stencil.
For read, \sys{} reads the requested metadata blocks from the secure storage and deposits it in the communication channel, which will be collected by the untrusted \fs{}; 
write is a mirror process.

4) A cloud server for running the metadata-only \fs{} replica.
We implement the server in a straightforward way, which keeps listening, parsing, and executing the file operations from the \device{} on the \fs{} replica.
Since the server must respond to smart devices with block operations, we run a privileged service on the server to extract the block operations to the user space.
Note that an alternative is to use userspace block device (\code{BUSE}~\cite{buse}), which does not require a privileged service on the cloud server. 



\paragraph{Smart device applications}
Atop \sys{}, we build three applications derived from real-world smart devices:

\begin{myenumerate}
	
	\item A \textbf{cleaning robot} (\code{Robot}) for house cleaning~\cite{rev-eng-mi-robot}.
	We derive the workloads from Xiaomi Vacuum Robot~\cite{vacuum-log}.
	When cleaning, the robot sequentially updates a log file every 200ms;
	each entry is 32-byte 3-tuple containing coordinates (x, y) and the angle. 
	After cleaning, the robot reads back the whole log file, reconstructs the cleaning map using SLAM in PPM format; each pixel of the map is 5cm in physical world and a typical map size is therefore tens to a few hundred KB.
	At last the robot writes back the map file to disk.
	We fix the log size to 512KB and map size to 256KB.

	
	\item A \textbf{voice assistant} (\code{Voice}) for interacting with user speech commands~\cite{mycroft}. 
	We derive the workloads from Mycroft~\cite{mycroft}.
	When starting, the voice assistant first loads a set of user-defined rules, called \textit{skills}, to respond to user speech;
	each skill is a directory under \code{\textasciitilde/.mycroft/} containing a \code{.voc} and a \code{.intent} file to identify the intent of user, a \code{.dialog} file to respond to user, and a \code{.json} file to describe the skill;
	every 2s the device will scan the three skill files using \code{fstat} in case any modification or updates.
	After boot, it listens to the user speech in the background, records the audio, and stores the recording as a \code{wav} file under \code{/tmp}; 
	it then runs speech to text transcription (STT) and responds to user according to the STT results and existing skill.

	\item An \textbf{intelligent camera} (\code{Camera}) for license plate recognition.
	We derive the workloads from intelligent traffic systems.
	As the surveillance goes on, the camera periodically captures and saves images (1080P in JPG format).
	Every other 10 seconds, it reads in the saved images and runs license plate detection algorithm on them;
	on the images, the algorithm detects canny edge, dilates them, and then find license plate blobs.
	Finally, it draws bounding box around detected plate blobs and saves the resulting image.
	We use images from UFPR-ALPR dataset~\cite{ufpr-alpr}.
	For the application, we ported to TEE \code{SOD}~\cite{libsod}, a popular embedded computer vision library.
	
\end{myenumerate}

	These applications serve as our macrobenchmarks. 

\section{Evaluation}
\label{sec:eval}

In this section, we seek to answer the following questions: 
\begin{myenumerate}
	\item How does \sys{} reduce TCB and thwart attacks?
	\item What is the security overhead of \sys{}?
	\item How does \sys{}'s API augmentation (\S\ref{sec:design:network}) impact programmability and reduce overhead?
\end{myenumerate}


\subsection{Security Analysis}
%
%

\begin{table}
\centering
	\vspace{1pt}
	\includegraphics[width=0.48\textwidth{}]{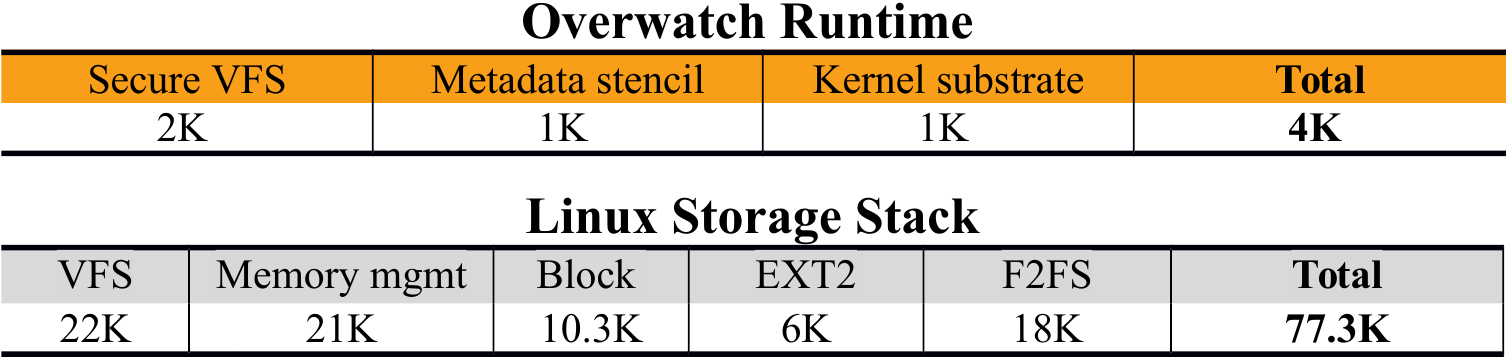}
	\caption{A comparison between the source of \sys{} and that of the Linux storage stack, showing that \sys{} reduces the on-device TCB significantly.}
	\label{tab:sloc}
\end{table}

\subsubsection{TCB analysis}

\paragraph{On-device TCB size} 
Table \ref{tab:sloc}(a) shows a breakdown of the \sys{} source code, which only adds 4K SLoC to the TCB. 
The size of the \sys{} binary is 52KB, a small fraction (3.3\%) of the entire OP-TEE binary. 

\paragraph{TCB interfaces}
The \sys{} secure runtime only exports two functions: one for issuing file API requests (to both \fs{} twins) and one for receiving block requests from the untrusted local \fs{}.
Two worlds share no state; 
all messages and the enclosing arguments are passed by value. 

\paragraph{Comparison to alternative TCB}
Compared to enclosing the entire Linux storage stack in TEE (the source count listed in Table~\ref{tab:sloc}(b), \sys{} significantly reduces the storage stack's on-device TCB by 16$\times$. 
This is because \sys{} completely excludes the file system logic (\code{ext2} and \code{f2fs} in our example) from the TEE and implements compact data functions for the TEE.


\subsubsection{Attacks thwarted by \sys{}}
\begin{table}[h]
	\fontsize{8.0}{9.0}\selectfont
	\begin{center}
		\setlength\tabcolsep{3pt}	
		\begin{tabular}{
				l	
				l	
				c	
				l 	
			}
			\hlineB{2.5}
			\textbf{Attack} &
			\textbf{Attack Vector} & 
			\textbf{Violated SG} &
			\textbf{Defense Mechanism} 
			\\ 
			\hlineB{1.5}
Iago attack				 	
& Block read
& \texttt{C---}
& Metadata stencil\\ 

Iago attack				 	
& Block write
& \texttt{-IFV}
& Operation Verifier\\ 

Iago attack				 	
& File descriptor
& \texttt{-IFV}
& Secure VFS\\ 

PF/Cache
& File \& offset
& \texttt{C---}
& TEE Property\\
%



\hlineB{2.5}
		\end{tabular}
		\fontsize{7.3}{7.7}\selectfont \\
		\textbf{SG}: security guarantees. \\ 
		\texttt{C}: user data confidentiality; 
		\texttt{I}: user data integrity;  
		\texttt{F}: user data freshness;
		\texttt{V}: verifiable persistence; 
		\vspace{2mm}
		\caption{List of attacks and defense mechanisms of \sys{}. PF/Cache denotes page fault and cache based attack.
		}
		\label{tab:sec-analysis}
	\end{center}
\end{table}

Table \ref{tab:sec-analysis} shows an overview of major attacks that target an \device{} to compromise our security guarantees.



\paragraph{Iago attack.}
As disclosed in~\cite{iago-attack}, a compromised kernel can forge return results of system services, subverting the efforts of prior approaches that delegate file operations to an untrusted \fs{}~\cite{sgx-fs,optee}.
For example, when prior approaches delegate \code{open("/ofs.txt")} to the untrusted \fs{}, a compromised \fs{} may request TEE for user data blocks instead of the only required metadata blocks of \code{"ofs.txt"}, hence resulting in  sensitive information leakage.

\sys{} defeats Iago attack in the following way:
first, the metadata stencil ensures only metadata can pass over to the untrusted \fs{} with any inlined user data erased.
Second, the operation verifier verifies each resultant user block request with the cloud, which ensures the legitimacy of user block requests.
Lastly, the user block data always stays in secure page cache and is never disclosed to the untrusted world.
Hence, with these components as a package, \sys{} defeats Iago attack in a straightforward way. 



\paragraph{Page fault \& cache side-channel attack.}
Page fault attacks are usually launched against SGX~\cite{os-side-channel}, since a compromised SGX driver is able to modify page attributes belong to SGX enclave and cause page faults to infer the memory access pattern.
However, by exploiting TrustZone, \sys{} is by design immune to such an attack because TEE memory is physically isolated and thus transparent to normal world.
Therefore, normal world is unable to tamper with TrustZone memory, and to launch page fault attacks.
To defeat cache side-channel attack, \sys{} fully confines cached read/write to Trustzone, and flushes its cache before switching back to the normal world.
As a result, no cache contention between secure world and normal world can occur to launch cache side-channel attack~\cite{trusense}, hence \sys{} thwarts cache side-channel attack.

\paragraph{What can be learnt by eavesdroppers?}
\begin{myitemize}
	\item The cloud server and the local OS, by running the twin \fs{}s, are able to access metadata which we set to reveal (\S\ref{sec:motiv:threats}).
	They access directory and file names in encrypted form but cannot decrypt them. 
	They observe file API invocations with API semantics only (\S\ref{sec:motiv:threats}).
	They observe uncached block access activities, while most of block accesses are cached in TEE and hence invisible.
	They cannot touch file data. 
	
		
	\item A network-level eavesdropper observes encrypted network traffic to/from the smart device. 
	She can infer the activities of file access and block access without knowing which files/blocks are accessed.
\end{myitemize}

\subsection{Methodology}
\label{sec:eval:bench}

\begin{table}
\centering
	\vspace{1pt}
	\includegraphics[width=0.48\textwidth{}]{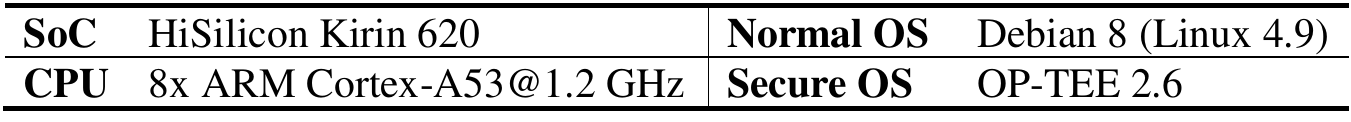}
	\caption{The test platform used as a smart device}
	\label{tab:plat}
\end{table}

\paragraph{Test setup}
We test \sys{} on Hikey~\cite{hikey}, an ARM-based development board, as the smart device. 
We choose this board for its good support for \tz{}. 
The details of Hikey are summarized in Table~\ref{tab:plat}. 
We run an x86 machine as the cloud server. 
The two machines are connected by Ethernet in an isolated LAN. 
We use Linux traffic control (\code{tc}) on the x64 machine to emulate different network conditions.

We test \sys{} with ext2 and f2fs, with two commodity, unmodified \fs{}s;
we choose them to represent different levels of complexity: 
while ext2 is classic and simple, f2fs is modern and feature-rich. 

We run a suite of macrobenchmarks and microbenchmarks. 
Prior to each run, we reboot both devices in order to have a cold cache.
As the current \tz{} TEE lacks device drivers for flash storage, 
we use ramdisk in TEE as the secure disk. 
We cap the ramdisk performance at 4K IOPS, which typical in today's low-cost flash. 

\paragraph{Macrobenchmarks}
To understand \sys{}'s end-to-end impact, we run the applications described in \sect{impl}. 
We define overhead metrics for applications based on their objectives. 
For Camera, we study its loss of throughput in image processing. 
For Voice and Robot, we study the extra delays they experience in each mission, i.e. data logging and map reconstruction, and responses to user voice commands respectively. 
We test under a spectrum of typical network delays according to recent study~\cite{4gperf}.

We show \sys{}'s impacts on latency of \code{Robot} and \code{Voice} as they are latency-sensitive.
We then shift the focus on its impacts on throughput of \code{Camera}.
This is because compared with \code{Robot} and \code{Voice} whose workloads are lightweight data logging, \code{Camera} is more compute-intensive and thus stresses throughput more. 


\paragraph{Microbenchmarks}
To understand the overhead of \sys{}'s on-device security mechanisms, we run a series of stress tests. 
We run Iozone v3.482~\cite{iozone}, a widely adopted \fs{} benchmark suites for Linux. 
We configure \sys{} to bypass validation with the cloud replica while still keeping all on-device mechanisms on.
To exclude the benefit of cache and observe the security overhead, we set the \code{O\_DIRECT} flag in file APIs to force each read/write hit the \sys{}.
We then compare \sys{} with native, insecure \fs{}s which run in the normal world atop ramdisk.


\begin{figure*}[t!]
	\centering
	\includegraphics[width=\textwidth{}]{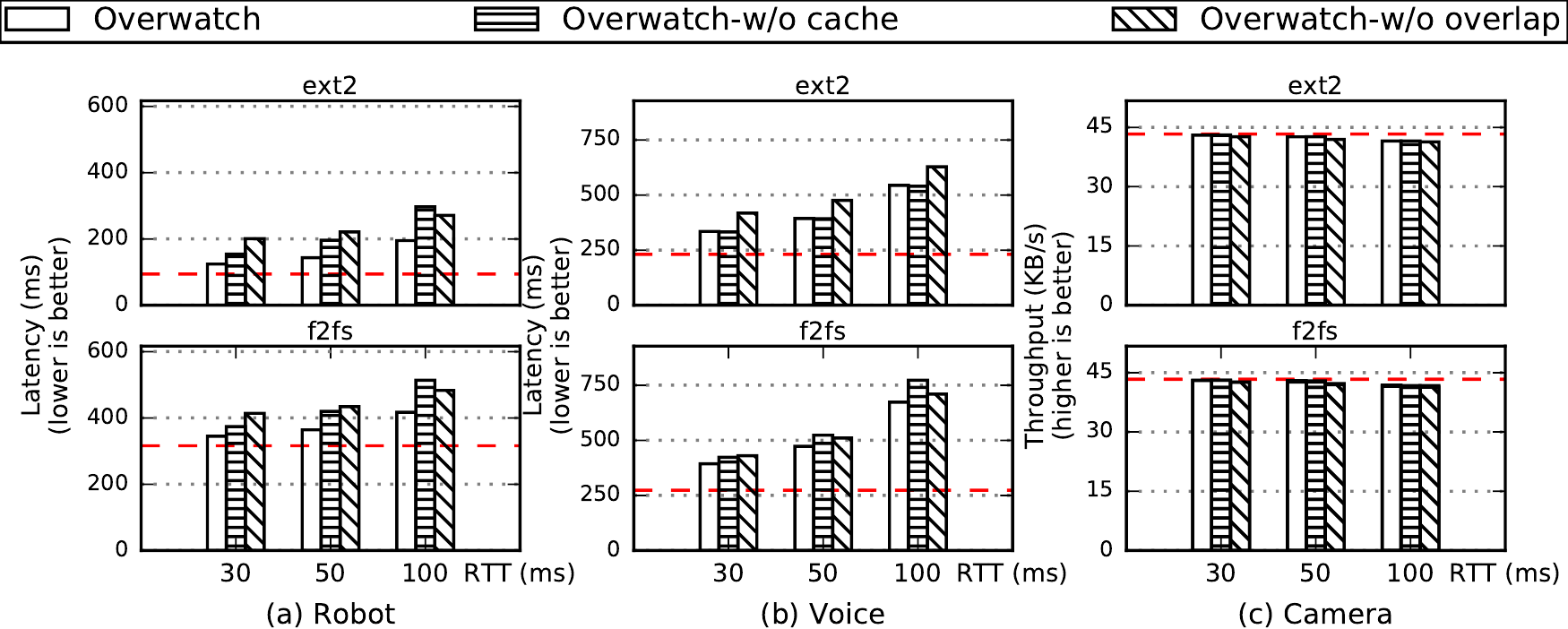}
	\caption{\sys{}'s impact on application performance under different  network latencies (x-axis). 
	Red dashed line: native performance when applications run on insecure \fs{}s. 
	The two additional \sys{} versions show secure cache and overlapping network/storage operations contribute to reduce the overhead.	
	}
	\label{fig:real}
\end{figure*}

\subsection{Application security overhead}
\label{sec:macrobenchmark}

Our results show \sys{} adds moderate overhead to the representative \device{} applications.

\paragraph{Application latency increase}
\sys{} adds moderate latency overhead to the application, considering different network delay and on-device security hardening.

Overall, at the common latency segment (50ms), \sys{} incurs as little overhead as 15\% in \code{Robot} (f2fs).
\sys{} achieves such low overhead due to the asynchronous writes -- the application does not block while waiting for the blocks to be verified. 
Moreover, the secure page cache facilitates trusted read without going to the cloud, it saves significant RTT in the \code{Robot} benchmark where the whole log is read back for map reconstruction.
In \code{Voice}, \sys{} incurs 45\% overhead.
This is because \code{Voice} requires three \code{fstat} in each run, which is sensitive to the network delay due to its synchronous nature. (\S\ref{sec:impl})

\paragraph{Application throughput loss}
As shown in Figure~\ref{fig:real}, in a broad latency spectrum, \sys{} causes the throughput of \code{Camera} to drop by no more than 5\%.
The low overhead is due to two factors: 
1) for read, the storage architecture which caches recently produced image data, reducing consultation with the cloud replica and the secure disk; 
2) for write, most operations are asynchronous and hence effectively overlap with the application's processing of subsequent images. 

\paragraph{Network bandwidth usage} \sys{} incurs light overhead in  network bandwidth usage, as the \device{} and the cloud only exchange compact validations instead of actual file data.
Even in Camera, our most data-intensive application, the smart device uses uplink/downlink bandwidth at 3.7 KB/sec and 2.6 KB/sec respectively, which are minor as compared to typical wireless bandwidth today (hundreds KB/sec)~\cite{4gperf}.


\paragraph{Cloud server overhead} 
In running the \fs{} replica, the cloud server sees negligible CPU overhead as the execution is bound by network delays. 
Possessing only metadata, the replica is also space-efficient.
To support a 4GB disk on device, the metadata stored by the cloud replica are up to 12MB for \code{f2fs}, and up to 65MB for \code{ext2}. 
This implies that every 1TB cloud storage server can support up to 87k and 16k smart device instances running \code{f2fs} and \code{ext2}, respectively.

\begin{figure*}[t!]
	\centering
	\includegraphics[width=\textwidth{}]{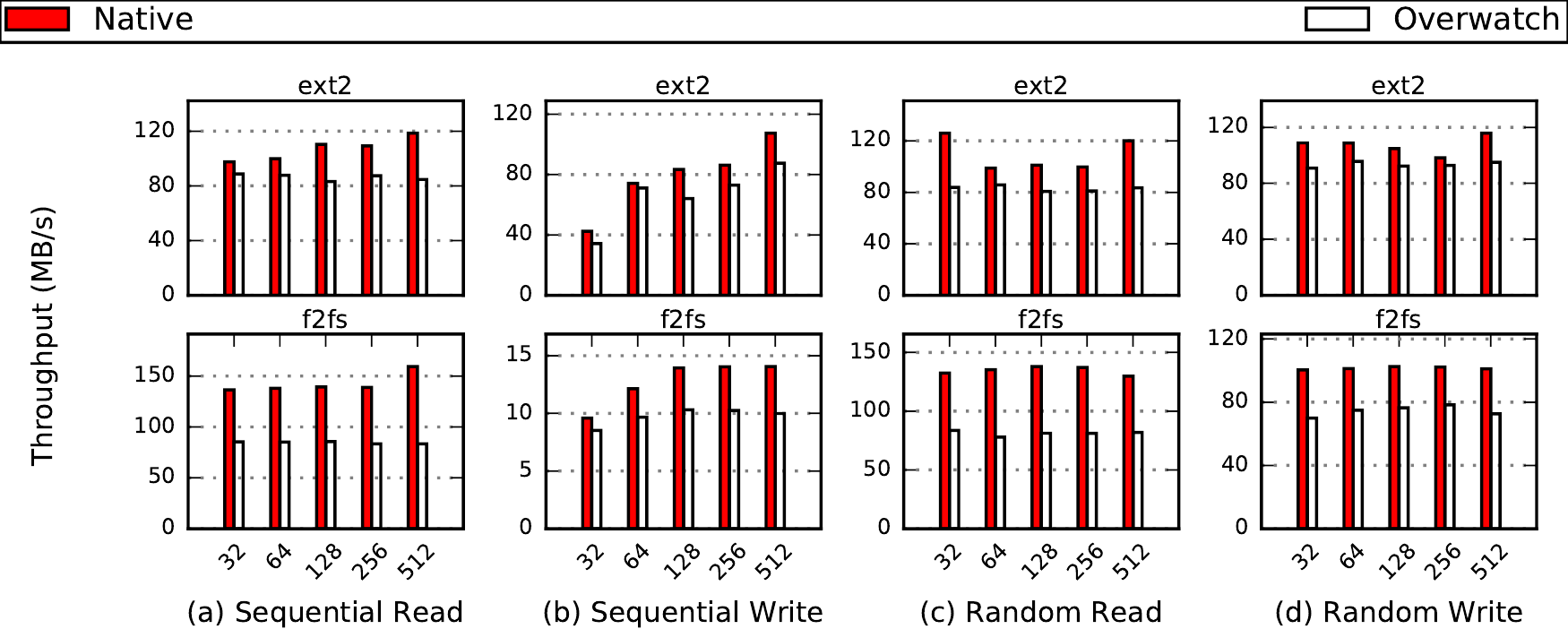}
	\caption{Microbenchmark performance of \sys{} compared with native, insecure \fs{}s. Benchmark: Iozone~\cite{iozone}. Read/write size = 4KB. O\_DIRECT flag set.}
	\label{fig:microbenchmarks}
\end{figure*}

\subsection{Security overhead under stress test}
\label{sec:microbenchmark}

We run the aforementioned stress test to understand the overhead of \sys{}'s on-device components. 


\paragraph{Overhead of \sys{} execution}
\sys{}'s on-device security mechanisms introduce noticeable overhead: because of them, the throughput of microbenchmark drops by 25\% on average (24\% for sequential and 26\% for random). 
Between the two commodity \fs{}s we tested, \code{f2fs} experiences higher overhead (33\% on average) than \code{ext2} (18\% on average), likely due to the former's more sophistic logic and hence more (sometimes 10x) block requests. 

\paragraph{Impact of metadata stencil.}
The metadata stencil examines each outgoing block served to the local \fs{}. 
For ext2 in which file data and metadata do not co-locate on the same block, checking a block against its metadata stencil is simply examining a flag (\S\ref{sec:design:stencil}). 
This incurs minor overhead, 
e.g. compared to \code{native-ext2}, the sequential read throughput of  \code{overwatch-ext2} drops by 18\%. 
However, for f2fs in which file data and metadata may co-locate, 
applying metadata stencil requires to redact inline file data in the outgoing block.  
This reduces throughput by up to 41\%, shown in the bottom figures of Figure~\ref{fig:microbenchmarks} (a) and (c).



\noindent
\textbf{Cross-world invocations} cause low delays. 
It only takes few thousand nanoseconds for the normal world \fs{} to receive file APIs sent by the TEE, and vice versa. 
In a file API (e.g. \code{fstat}) that must invoke both local/cloud \fs{}s synchronously, the cross-world invocation delay is negligible as compared to the network delay (in milliseconds). 
Within a cross-world invocation, the world switch takes around 49ns.

\begin{figure}[h]
	\centering
	\includegraphics[width=0.4\textwidth{}]{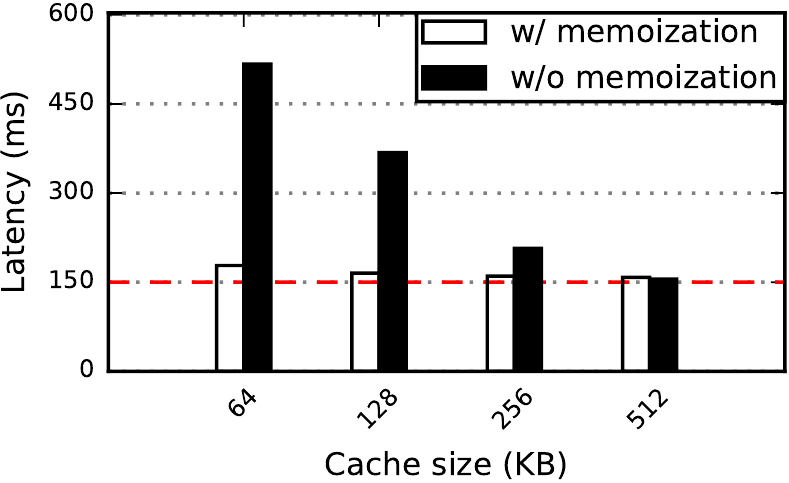}
	\vspace{-3mm}
	\caption{Impact of different sizes of secure page cache with and without memoizing validations. File size is 512KB and the network RTT is 50ms. Dashed line shows native (insecure) performance.}
	\label{fig:memoiz}
	\vspace{-4mm}
\end{figure}

\noindent
\textbf{Memoizing validations} effectively reduces network trips and hence overhead. 
We demonstrate this with the benchmark, which writes 512KB to the file and read them back for processing;
this is similar to Robot benchmark and the write-read back for processing pattern can be commonly found in other edge processing devices. 
As shown in Figure~\ref{fig:memoiz}, memoizing reduces latency significantly, by up to 291\%. 
The saving is more pronounced with smaller secure page cache.  
This is because even an access to a file offset misses the page cache and has to hit the secure disk, 
\sys{} still knows which block the offset maps to and can safely skip consulting local/cloud \fs{}s. 
The saving is substantial because consulting the cloud \fs{} is much slower than accessing the secure disk. 
\begin{figure}[t]
	\centering
	\includegraphics[width=0.46\textwidth{}]{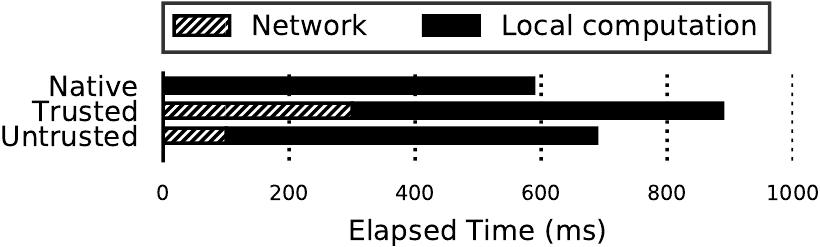}
	~ 
	\vspace{-5pt}		
	\caption{Application latency of the \textbf{Camera} app, showing the comparison among native (insecure), the  execution time with that of adopting \code{UNTRUST} and \code{TRUSTED} api. Network RTT is 100ms.}
	\label{fig:app}
	\vspace{-3mm}		
\end{figure}

\vspace*{-4mm}
\subsection{A case study of using the augmented API}


\definecolor{codegreen}{rgb}{0,0.6,0}
\definecolor{codegray}{rgb}{0.5,0.5,0.5}
\definecolor{codepurple}{rgb}{0.58,0,0.82}
\definecolor{backcolour}{rgb}{0.95,0.95,0.92}

\lstset{emph={%
	VALIDATION, O_UNTRUSTED%
	},emphstyle={\color{red}\bfseries}%
}%

\lstdefinestyle{mystyle}{
	commentstyle=\color{codegreen},
	keywordstyle=\color{magenta},
	numberstyle=\color{codegray},
	stringstyle=\color{codepurple},
	basicstyle=\fontsize{9}{9}\selectfont\ttfamily,
	breakatwhitespace=false,         
	breaklines=true,                 
	captionpos=b,                    
	keepspaces=true,                 
	numbers=left,                    
	showspaces=false,                
	showstringspaces=false,
	showtabs=false,                  
	tabsize=2
}

\lstset{style=mystyle}
\lstset{
xleftmargin=5.0ex,
framexleftmargin=5.0ex,
frame=tb, 
breaklines=true,
numberblanklines=false,
captionpos=b,
numbers=left,
label=list:api}
\begin{lstlisting}[caption= Code example showing the use of untrusted reads. The augmented API is highlighted in bold and red. ,escapechar=!,label={code:api}]
for (i = 0; i < N_FILES; i++) {
	/*Allow read to return untrusted data*/
	fd_in=open(in[i],O_UNTRUSTED|O_RDONLY);!\label{line:open}!
	fd_out=open(out[i],O_CREAT|O_WRONLY); !\label{line:open2}!

	/*Determine data size for read*/
	fstat(fd_in, &stat);
	/*Read untrusted data*/
	ret=read(fd_in,buf,stat.st_size); !\label{line:read}!

	/* Make image */
	img = make_image(buf, IMG_JPG);
	/* Processing untrusted data. slow. */
	ret = detect_license(img, buf, &sz); 

	/*Wait for validation*/
	ret = select(fd_in, VALIDATION, ...); !\label{line:select}!
	if (ret == SUCCESS) { /* validated */
		/* Write results to new file */
		write(fd_out, buf, sz); !\label{line:write}!
	} else { /* validation failed */
	   /*..discard compute results..*/
	}
	close(fd_in);	
	close(fd_out);
}
\end{lstlisting}

\sys{} gives applications an option to access file data before validation arrives (\S\ref{sec:design:memoiz}) and hence hide network latency. 
To understand the entailed programming burden and performance reward, we build a revision of the camera application. 
The application open a series of input image files, runs license detection on them, and write results to output image files. 
In the new version, the application opens input images files by specifying the \textit{untrusted} flag (line~\ref{line:open},~\ref{line:open2}).
When the application reads in images, \sys{} therefore returns the data as soon as the local \fs{} gives the block numbers.
As the application gets the untrusted image data, it computes, and executes a validation barrier (line~\ref{line:select}).
Only on successful validation (i.e. the vision algorithm has executed on a correct input image) will the application write back the processed image (line~\ref{line:write}).
With minor source changes, the application overlaps the vision compute with network delay in validating a \code{read}.

Figure~\ref{fig:app} shows the application latency reduction as the reward from handling the untrusted data. 
With fully trusted read/write, the application incurs 26\% higher latency as compared to a native execution baseline; 
with untrusted read, that overhead is reduced to 8\%. 
Note that, the use of untrusted file access is optional.

\vspace{-3mm}
\section{Related Work}
\label{sec:related}


We next discuss related work that is not covered so far.

\paragraph{Defending against untrusted OSes with TEE} 
Recent research has recognized OSes as untrusted~\cite{iago-attack} and examined various countermeasures against them, e.g. by relying trusted hypervisors~\cite{overshadow, trustvisor, inktag} and trusted compilers~\cite{virtualghost}.
To protect important software components against untrusted OSes, many systems use TEEs.
Haven~\cite{haven} protects unmodified apps and ports \code{FAT32} into an SGX enclave. 
Scone~\cite{scone} secures containers with TEE. Graphene~\cite{graphene-sgx} ports a library OS to protect apps in TEE. Obliviate~\cite{obliviate} exploits ORAM to keep file operations in TEE oblivious.
TrustShadow~\cite{trustshadow} focuses on protecting memory integrity of apps, and delegates syscalls to an untrusted OS.
Similar to them, we do not trust the OSes, and use TEE to shield critical execution; 
unlike them, we focus on \fs{} and propose new partitioning between data functions and control functions for protection; we also support multiple commodity \fs{}s.

Much work uses TEEs (mostly \tz{}) to enforce system software invariants. 
TZ-RKP~\cite{tzrkp} intercepts and examines control-critical instructions (e.g., pagetable update);
Sprobes~\cite{sprobes} checks kernel integrity to guard its code integrity.
Like much of them, we build on TrustZone; 
different from them, we do not enforce any invariant for ensuring integrity but instead verifying with the cloud.
Trustgyges~\cite{trustgyges} exploits TEE to hide file data from an untrusted smartphone OS. 
It neither partitions the storage stack nor use the cloud for validation.

\paragraph{Replicated execution.} 
The idea of replicated execution traces back to 1968~\cite{replicated-exectuion-debugging} and is used to detect and locate bugs, provide fault-tolerance~\cite{eve,state-machine-repl,rex}, and offer security guarantees~\cite{n-variant, diehard,repl-exec-web}.
Compared to them, we adopt the idea of replicated execution for security, but are the first to apply it to \fs{}s to our knowledge.
Replicated execution is also used for resource efficiency~\cite{remote-exec}.
Tango~\cite{tango}, MAUI~\cite{maui}, and Clonecloud~\cite{clonecloud} offload compute-intensive code from smartphones to the cloud.
Like them, we exploit the collaboration of cloud and the local device;
unlike them, we focus on using the cloud for security without disclosing user data. 
Knockoff~\cite{knockoff} ships IO traces from clients to the cloud for  creating multiple versions of disk image.
Similar to it, \sys{} exchanges file/block requests with the cloud;
unlike it, \sys{} does not upload file data but use the cloud as the verifier for \fs{} behaviors. 


\paragraph{Trustworthy file systems.}
A traditional approach to keeping file data confidential is through cryptographic \fs{}, either stackable~\cite{cfs,ncryptfs} or integrated natively~\cite{ecryptfs}. 
Compared to them, we do not trust the commodity kernel but reuse its \fs{} logic. 
Cryptographic \fs{}s are also designed for distributed computing. 
Plutus~\cite{plutus} supports file-sharing on untrusted storage with cryptographic key distribution and management.
SiRius~\cite{sirius} further guarantees freshness using Merkle tree.
SUNDR~\cite{sundr} detects tamper on files on untrusted servers.
While such prior work and \sys{} share the goal of file system security for networked computers, \sys{} offers strong security guarantees, e.g. assured persistence, and therefore introduces different techniques. 
\vspace{-3mm}
\section{Conclusions}
\label{sec:conclusion}
Smart devices produce security-critical data and keep them persistent on local storage. 
The current storage software on smart devices offer weak security guarantees. 
We propose a new storage architecture to provide data confidentiality, integrity, and assured persistence. 
Following an outsource-and-verify approach, we isolate file data in TEE, shielding it from commodity \fs{}s. 
We execute a metadata-only replica of the \fs{} in the cloud, which continuously validate the behaviors of the on-device \fs{}. 
We implement \sys{} and analyze its security properties. 
We demonstrate that \sys{} works with commodity file system and incurs moderate overhead. 
\sys{} represents a new design point for storage stacks.

	

	\scriptsize

	\bibliographystyle{bib/abbrv-minimal}

	
	
\end{document}